\documentclass[9pt,shortpaper,twoside,web]{ieeetran}
\usepackage{cite}
\usepackage{amsmath,amssymb,amsfonts}
\usepackage{graphicx}

\usepackage{siunitx}
\usepackage{color}
\usepackage{upgreek}
\usepackage{mathtools}
\usepackage{hyperref}

\begin{document}
\title{Revised Theoretical Limit of Subthreshold Swing\\in Field-Effect Transistors}

\author{Arnout Beckers, Farzan Jazaeri, and Christian Enz, \IEEEmembership{Fellow, IEEE}
	
	\thanks{This project has received funding from the European Union's Horizon 2020 Research \& Innovation Programme under grant agreement No. 688539 MOS-Quito (MOS-based Quantum Information Technology).}
	\thanks{The authors are with the Integrated Circuits Laboratory (ICLAB) at the Ecole Polytechnique F\'ed\'erale de Lausanne (EPFL), Microcity, 2000 Neuch\^atel, Switzerland (e-mail: arnout.beckers@epfl.ch).}
	
}

\maketitle

\begin{abstract}
This letter reports a temperature-dependent limit for the subthreshold swing in MOSFETs that deviates from the Boltzmann limit at deep-cryogenic temperatures. Below a critical  temperature, the derived limit saturates to a value that is independent of temperature and  proportional to the extent of a band tail. Since the saturation is universally observed in different types of MOSFETs (regardless of dimension or semiconductor material), the band tail is attributed to the finite periodicity of the lattice in a semiconductor volume, and to a lesser extent to additional lattice perturbations such as defects or disorder. 
\end{abstract}


\section{Introduction}
\begin{figure}[t]
	\centering
	\includegraphics[width=0.45\textwidth]{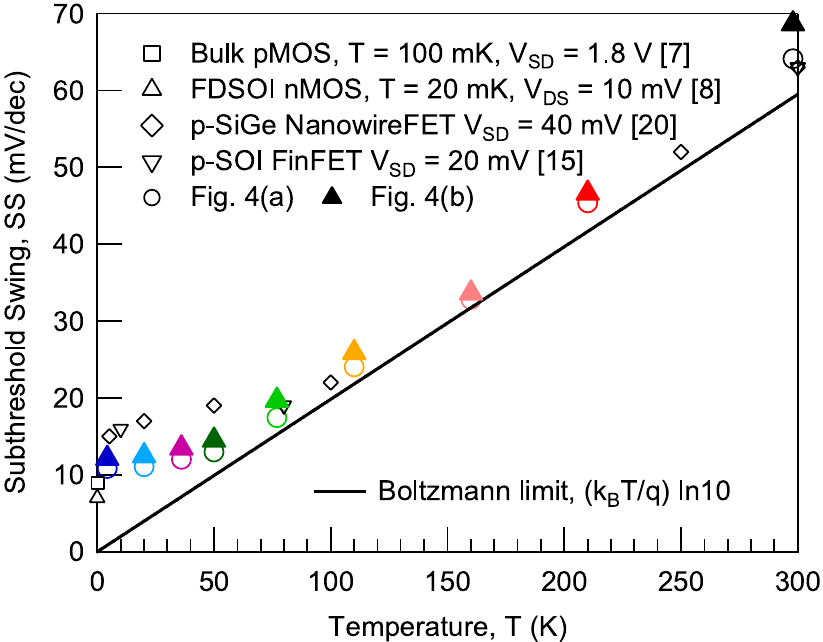}
	\caption{Saturating $SS(T)$ measured in different FET technologies deviating from the Boltzmann limit. Colored markers are obtained from our measurements in Figs.\,\ref{fig:meas}(a) and \ref{fig:meas}(b) at $I_{DS}\!=\!\SI{100}{\pico\ampere}$ and $\SI{1}{\nano\ampere}$, respectively. All devices have gate lengths in the $\si{\micro\meter}$-range.}
	\label{fig:overview}
\end{figure}

The Boltzmann limit of the subthreshold swing in FETs, $SS=(k_BT/q)\ln10$, predicts at room temperature the well-known $\approx$ \SI{60}{\milli\volt}/dec, and at deep-cryogenic temperatures ($< \approx$ \SI{50}{\kelvin}) an almost ideal, step-like switch ($k_BT/q$ is the thermal voltage). However, the measurements in FETs at deep-cryogenic temperatures reach merely $\approx$ 11 instead of \SI{0.8}{\milli\volt}/dec at \SI{4.2}{\kelvin}~\cite{kamgar_subthreshold_1982,hafez_assessment_1990,balestra_physics_2017,beckers_essderc,beckers_jeds,harald}, $\approx$ \SI{9}{\milli\volt}/dec instead of \SI{20}{\micro\volt}/dec at \SI{100}{\milli\kelvin}~\cite{incandela}, and $\approx$ \SI{7}{\milli\volt}/dec instead of \SI{4}{\micro\volt}/dec at \SI{20}{\milli\kelvin}~\cite{galy}. As shown in Fig.\,\ref{fig:overview}, this degradation is measured in structurally different FETs, operating in subthreshold at both low and high drain voltage ($V_{DS}$) and for various technologies:  
mature and advanced bulk and FDSOI MOSFETs \cite{gutierrez2000low,balestra2001device,beckers_essderc,beckers_jeds,incandela,harald,elewa_performance_1990,shin_low_2014,bohus_snw,solidstate,galy}, FinFETs \cite{achour_dc_2013,cretu_assessment_2016}, gate-all-around Si nanowire FETs \cite{boudier}, junctionless FETs \cite{trevisoli_effect_2014,trevisoli_junctionless_2016}, SiGe FETs \cite{paz_cryogenic_2018}, InP HEMTs \cite{schleeh}, SiC FETs \cite{kobayashi_interface_2016}, etc. Figure \ref{fig:overview} highlights this measured trend, deviating from the Boltzmann limit below a critical temperature, and then saturating to a value depending on the technology. The difference between the measured $SS$ and the Boltzmann limit is referred to as excess $SS$. The additional power that the FET consumes at deep-cryogenic temperatures due to the excess $SS$ is a crucial metric for the realization of quantum processors in silicon~\cite{pla2012single,morton2011embracing,maurand_cmos_2016,vandersypen2017interfacing,charb,schaal2019cmos,tedpaper,farzan} and for assessing the benefits of temperature scaling as an alternative to traditional scaling \cite{jamaldeen,gutierrez2000low,balestra2001device}. 

It is simply not possible to explain this saturation of $SS$ using the Boltzmann limit. Indeed, the Boltzmann limit is linear in $T$, and its slope versus $T$ is proportional to the slope factor ($m_0=1+C_{depl}/C_{ox}$) which is limited to 2 when neglecting the interface traps since $C_{depl}<C_{ox}$ ($C_{ox}$ is the gate-oxide capacitance, and $C_{depl}$ the depletion capacitance). Assuming a uniform density of interface traps over energy in the bandgap, does not help to model the behavior below \SI{50}{\kelvin}, since it only further increases the linear slope of $SS$ versus $T$ ($m=m_0+qN_{it}/C_{ox}$ where $N_{it}$ is the number of interface states per unit area). Furthermore, this approach has led to unreasonably high $N_{it}$ at deep-cryogenic temperatures. Typical $N_{it}$ values that have been reported in the literature are in the order of $\SI{e13}{}-\SI{e14}{\per\centi\meter\squared}$ at \SI{4.2}{\kelvin}~\cite{trevisoli_effect_2014,trevisoli_junctionless_2016,cretu_assessment_2016}, and \SI{e16}{\per\centi\meter\squared} at \SI{20}{mK}~\cite{galy}. The values at \SI{4.2}{\kelvin} are still possible in principle. The values at \SI{20}{\milli\kelvin}, however, exceed $\SI{7e14}{\per\centi\meter\squared}$ corresponding to the number of atomic lattice sites per unit area in silicon. Furthermore, it should be emphasized that the Boltzmann limit leads to a singularity in $N_{it}$ near \SI{0}{\kelvin}. Recently, relying on numerical simulations Bohuslavskyi et al. demonstrated  that an exponential band tail and Fermi-Dirac statistics leads to saturation of $SS$ at deep-cryogenic temperatures \cite{bohus,phdthesisheorhii}. The presence of a band tail in FDSOI FETs was explained by a combination of crystalline disorder, strain, residual impurities, etc. However, an imperfect band edge can already develop in a piece of semiconductor that is free from disorder or defects, but not infinitely periodic. Indeed, periodic boundary conditions are usually assumed for 3-D density-of-states ($DOS$) calculations which result in a perfectly sharp edge of the conduction band. Similarly for an electron in a 1-D periodic potential, the band edges are only perfect when the potential is infinitely periodic (Kr\"{o}nig-Penney model). Invoking the finite periodicity of the crystal in a MOS device could give a better explanation why the saturation of $SS$ is so universal among different MOS technologies. The saturation has been measured in older technologies as well, before strain and nanometer dimensions were introduced that lead to disorder. Furthermore, little statistical variation on $SS(\SI{4.2}{\kelvin})$ has been reported for 50 samples of the same technology (28-nm FDSOI)\cite{phdthesisheorhii}. While defects and disorder vary among devices, all devices on a wafer have a similarly broken periodicity in the direction of the MOS interface due to wafer cleavage followed by lattice-matched material growth. The little statistical variation is then due to other lattice perturbations such as the ones proposed by Bohuslavskyi et al.~\cite{bohus,phdthesisheorhii}. This is consistent with the fact that the $SS$ in FDSOI devices improves when the channel is displaced away from the front-gate interface by back-gate biasing~\cite{bohus}.\newpage

\section{\label{sec:deriv}Revised Theoretical Limit}
\begin{figure}[t]
	\centering
	\includegraphics[width=0.4\textwidth]{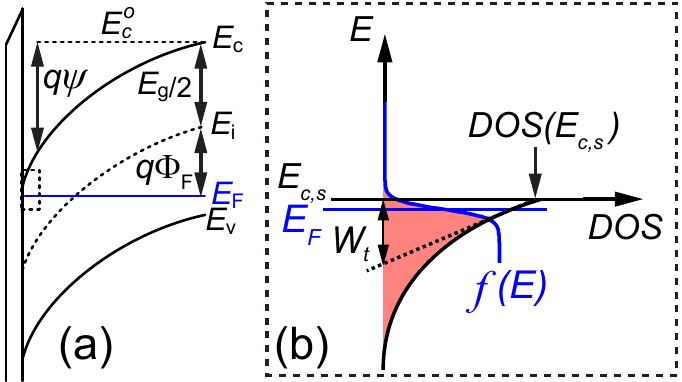}
	\caption{(a) MOSFET band diagram with (b) a zoom-in on the band tail at the surface. $W_t$ denotes the characteristic width of the exponential band tail, and $\psi\triangleq(E_c^o-E_c)/q$. Red area indicates the electron density.} 
	\label{fig:band}
\end{figure}

\begin{figure}[t]
	\centering
	\includegraphics[width=0.45\textwidth]{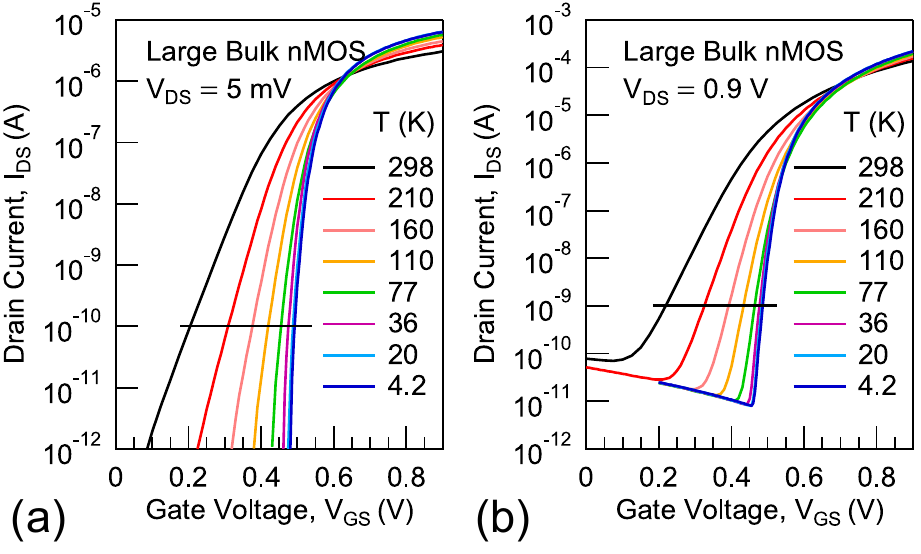}
	\caption{Measured transfer characteristics down to \SI{4.2}{\kelvin} in an $n$-channel, bulk-Si MOSFET with $W/L=\SI{3}{\micro\meter}/\SI{1}{\micro\meter}$ from a commercial 28-nm bulk CMOS process, a) at low $V_{DS}$ and  b) high $V_{DS}$. Horizontal lines indicate current levels at which $SS$ was extracted and shown in Fig.\,\ref{fig:overview}.}
	\label{fig:meas}
\end{figure}

The total drain current in subthreshold can be approximated by $I_{DS}=q(W/L)\mu(k_BT/q)(n_D-n_S)$, assuming a standard, bulk $n$-channel MOSFET, where $q$ is the electron charge, $W/L$ the width-over-length ratio of the transistor gate, $\mu$ the free-carrier mobility (assumed constant along the channel), and $n_D$ and $n_S$ the electron densities at the drain and source sides~\cite{taur}. Hence, $SS=\partial V_{GS}/\partial \log I_{DS}$ can be expressed as $m[(n_D-n_S)/(\partial n_D/\partial \psi_s-\partial n_S/\partial \psi_s)]\ln10$, where $V_{GS}$ is the gate-to-source voltage, $m=\partial V_{GS}/\partial \psi_s$ is the slope factor, and $\psi_s$ is the electrostatic potential at the surface compared to the bulk [Fig.\,\ref{fig:band}(a)]. We assume that $m=1+(C_{depl}+C_{it})/C_{ox}$ where $C_{it}$ is the interface-trap capacitance. The electron density in the band tail [Fig.\,\ref{fig:band}(b)] is described by: \begin{equation}\label{eq:int}
n=\int_{-\infty}^{E_{c,s}}DOS(E_{c,s})\exp\left(\frac{E-E_{c,s}}{W_t}\right)f(E)dE,
\end{equation}
where $E_{c,s}$ is the conduction-band energy of the sharp band edge at the surface, $W_t$ is the characteristic decay of an exponential band tail in the bandgap, and $f(E)$ is the Fermi-Dirac function. For simplicity, since $SS$ will not depend on the exact value of $DOS(E_{c,s})$, we assume that $DOS(E_{c,s})$ can be given by the  conduction-band $DOS$ in 2-D:  $N_c^{2D}=g_vm^*/(\pi\hbar^2)$, where $g_v=2$ is the degeneracy factor, $m^*=0.19\,m_e$ is the effective mass in silicon (assumed temperature independent), $m_e$ the electron mass, and $\hbar$ the reduced Planck constant. The solution of integral (\ref{eq:int}) takes the form of a Gaussian hypergeometric function ($\prescript{}{2}{F}_1=F_1$)~\cite{andrews1992special}:
\begin{equation}\label{eq:n}
n=N_c^{\mathrm{2D}}W_t F_1\left(1,\theta;\theta+1;z\right),
\end{equation}
where $\theta=k_BT/W_t$,  $z=-\exp\left[\left(E_{c,s}-E_{F,n}\right)/(k_BT)\right]$ and  $E_{F,n}=E_F-qV$ is the quasi-Fermi energy of electrons and $V$ is the channel voltage. The band diagram in Fig.\,\ref{fig:band}(a) shows that $E_{F}=E_c^o-E_g/2-q\Phi_\mathrm{F}$, where $E_c^o$ is the conduction-band energy in thermal equilibrium, $E_g$ the bandgap, and $\Phi_\mathrm{F}=(k_BT/q)\ln(N_A/n_i)$ the Fermi potential with $N_A$ the doping concentration and $n_i$ the intrinsic carrier concentration. Using $\psi_s\triangleq-(E_{c,s}-E_{c}^{o})/q$, it follows that $E_{F,n}-E_{c,s}=q\psi_s-E_g/2-q\Phi_\mathrm{F}-qV$. The latter can be inserted in (\ref{eq:n}) to yield $n$ as a function of $\psi_s$ where $z=-\exp\left[-q\psi_s^\prime/(k_BT)\right]$, $\psi_s^\prime=\psi_s-\psi^*_s$, and $\psi_s^*=E_g/(2q)+\Phi_\mathrm{F}+V$. The defined $\psi_s^*$ depends only on $T$ and $N_A$ at a fixed $V_{DS}$. Note that for $\psi_s$ in subthreshold, ranging from 0 (flatband) to $2\Phi_\mathrm{F}+V$ (threshold), $\psi_s^\prime$ is always negative. The first derivative of a hypergeometric function $F_1(a,b;c;z)$ is given by $(ab/c)F_1(a+1,b+1;c+1;z)$~\cite{abramowitz}. Differentiating (\ref{eq:n}) with respect to $\psi_s$ (applying the chain rule for $z$), we find that 
\begin{figure}[t]
	\centering
	\includegraphics[width=0.45\textwidth]{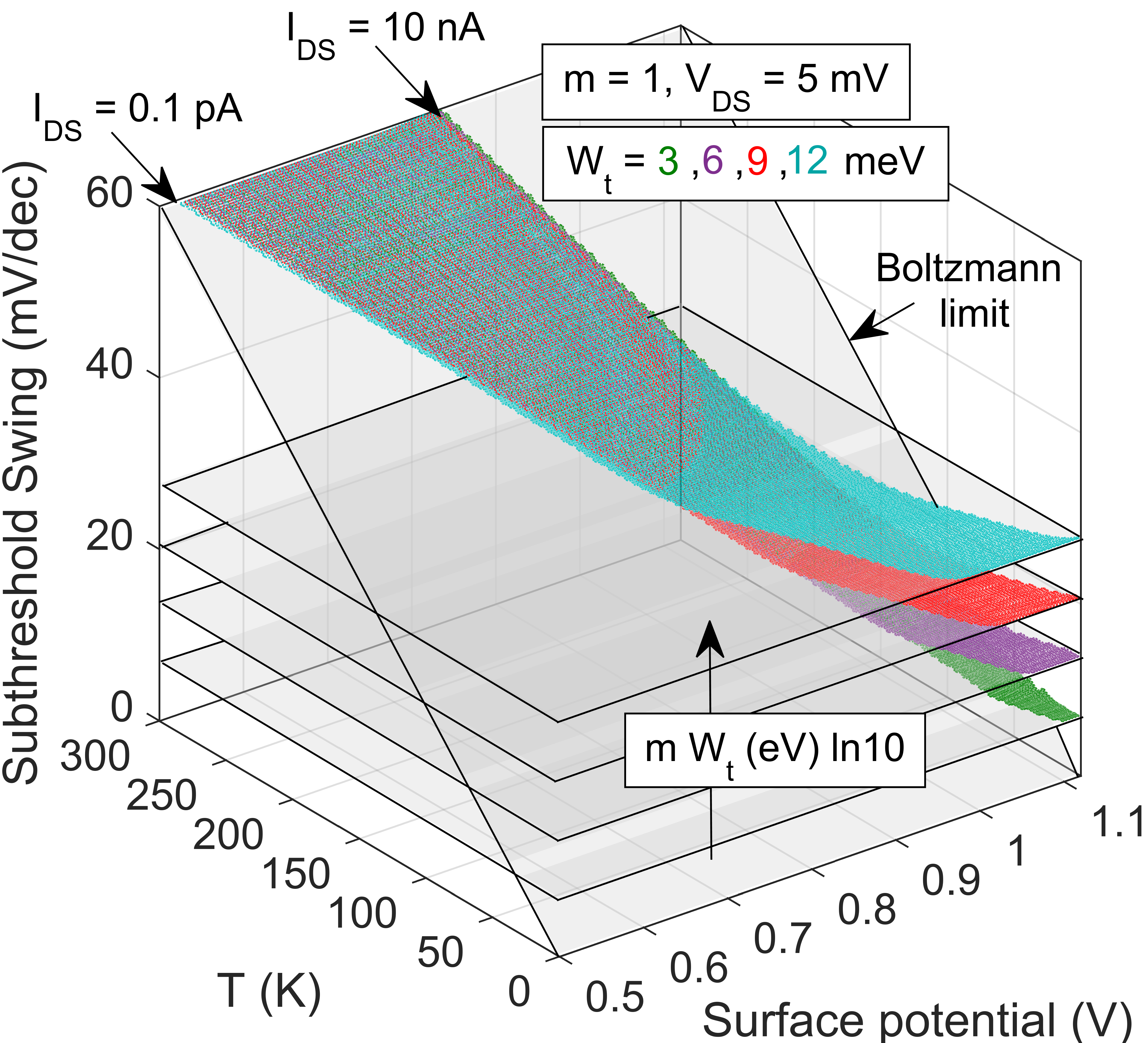}
	\caption{Revised limit of $SS$ (\ref{eq:ss})-(\ref{eq:A}) plotted for different $W_t$ (colored surfaces). The range of $\psi_s$ is limited to weak inversion (taken between the constant current values $\SI{0.1}{\pico\ampere}$ and $\SI{10}{\nano\ampere}$. A linear increase in the electron mobility is assumed from \SI{200}{\centi\meter\squared\per\volt\per\second} at \SI{300}{\kelvin} to \SI{800}{\centi\meter\squared\per\volt\per\second} at \SI{1}{\kelvin}~\cite{beckers_essderc}, $W/L=3$, and $N_A=\SI{e17}{\per\centi\meter\cubed}$.} 
	\label{fig:3d}
\end{figure}
\begin{eqnarray}
\frac{\partial n}{\partial \psi_s}=-qzN_c^{\mathrm{2D}}\frac{F_1\left(2,\theta+1;\theta+2;z \right)}{\theta +1}.
\label{eq:nderiv}
\end{eqnarray}
Inserting (\ref{eq:n}) and (\ref{eq:nderiv}) in the expression for $SS$, gives: 
\begin{equation}
SS=m\left(\frac{k_BT}{q}\right)\ln10\times A\left[z(\psi_s,V),T,W_t\right],
\label{eq:ss}
\end{equation}
where $A$ is given by
\begin{equation}
\frac{\left[\theta^{-1}F_1\left(1,\theta;\theta+1;z\right)\right]^{V=V_{DS}}_{V=0}}{\left[-z(\theta+1)^{-1}F_1\left(2,\theta+1;\theta+2;z\right)\right]^{V=V_{DS}}_{V=0}}.
\label{eq:A}
\end{equation}
Expression (\ref{eq:ss})-(\ref{eq:A}) is plotted in Fig.\,\ref{fig:3d} versus $T$ and $\psi_s$ for different $W_t$ together with the Boltzmann limit. As shown in Fig. \ref{fig:3d}, (i) $SS$ rolls off from the Boltzmann limit and saturates at deep-cryogenic temperatures, (ii) the saturation value of $SS$ increases with $W_t$, (iii) the critical temperature at which $SS$ starts to deviate from the Boltzmann limit increases with $W_t$. Figures~\ref{fig:meas}(a) and \ref{fig:meas}(b) show the measured transfer current characteristics in a large bulk silicon, $n$-channel MOSFET from \SI{298}{\kelvin} down to \SI{4.2}{\kelvin} and biased at low and high $V_{DS}$, respectively. The device was fabricated in a commercial 28-nm bulk CMOS process. The measurements were performed using a Lakeshore CPX cryogenic probe station and a Keysight B1500a semiconductor device analyzer. In Fig.\,\ref{fig:comp}, the derived limit is successfully compared with the measurements from Fig.\,\ref{fig:meas}(a) at low $V_{DS}$. The extracted value of $W_t$ is in agreement with the electron-spin-resonance measurements in silicon FETs at sub-Kelvin temperatures by Jock et al\cite{jock}.

\section{Saturation Value}
\begin{figure}[t]
	\centering
	\includegraphics[width=0.5\textwidth]{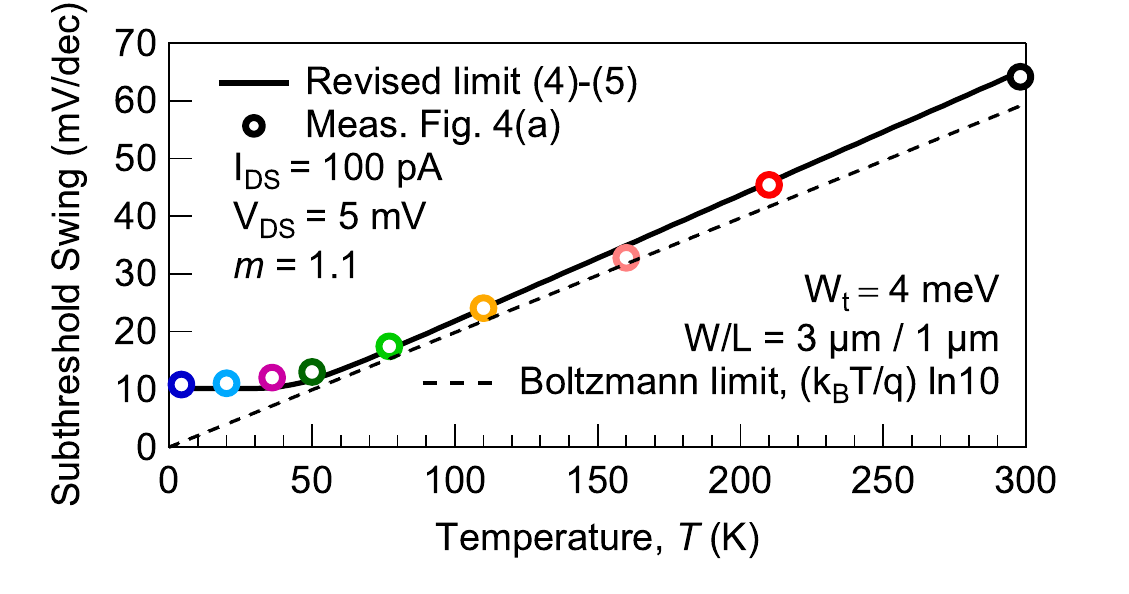}
	\vspace{-0.4cm}
	\caption{Experimental validation of the revised $SS$ limit from (\ref{eq:ss})-(\ref{eq:A}).} 
	\label{fig:comp}
\end{figure}

An expression for the saturation value of $SS$ at deep cryogenic temperatures ($k_BT\ll W_t$ or $\theta \rightarrow 0$) can be derived from (\ref{eq:ss})-(\ref{eq:A}):
\begin{equation}
SS^{k_BT\ll W_t}=m\left(\frac{W_t}{q}\right)\ln10\times\frac{F_1(1,0;1;z)}{F_1(2,1;2;z)}(-z)^{-1}
\label{eq:hyp}
\end{equation}
where $W_t$ is in Joules. Applying one of Euler's linear transformations for hypergeometric functions, i.e., $F_1(a,b;c;z)=(1-z)^{-b}F_1(b,c-a;c;z^\prime)$ \cite{abramowitz}, where $\vert z^\prime\vert=\vert z/(z-1)\vert < 1$ and $c=a$, gives 
\begin{equation}
SS^{k_BT\ll W_t}\!=\!m\left(\frac{W_t}{q}\right)\ln10\times\frac{F_1(0,0;1;z^\prime)}{F_1(1,0;2;z^\prime)}(1-z^{-1})
\label{eq:sssat}
\end{equation}
The hypergeometric functions in the numerator and denominator of (\ref{eq:sssat}) are both equal to one, because either $a$ or $b$ is zero in the series representation \cite{abramowitz}. Since $\psi_s^\prime$ is negative and $k_BT$ small, $z^{-1}\rightarrow 0$, we obtain a saturation value that is independent of $T$, $\psi_s$, and $V_{DS}$, and proportional to $W_t$: 
\begin{equation}
SS^{k_BT\ll W_t}=m\left(\frac{W_t}{q}\right)\ln10
\label{eq:satval}
\end{equation}
The above result confirms the saturation value of $SS$ that was apparent in the numerical simulations by Bohuslavskyi et al.~\cite{bohus,phdthesisheorhii}. The revised limit in (\ref{eq:ss})-(\ref{eq:A}) and saturation value in (\ref{eq:satval}) are valid for bulk, SOI, FinFET, nanowire, and other multigate FETs, provided that $m$ accounts for enhanced electrostatic control. The horizontal planes in Fig.\,\ref{fig:3d} indicate the saturation values for increasing $W_t$. The critical temperature for which $SS$ deviates from the Boltzmann limit can be estimated as $T_{crit}=W_t/k_B$. In Fig.\,\ref{fig:comp}, $T_{crit}$ is about \SI{46}{\kelvin}. Similarly, $T_{crit}$ provides a simple method to obtain $W_t$ from dc measurements by plotting $SS$ versus $T$. Furthermore, (\ref{eq:satval}) demonstrates that the deep-cryogenic subthreshold performance of FETs is determined by $W_t$ and $N_{it}$ (in $m$). A more reasonable $N_{it}$ can be extracted at sub-Kelvin temperatures by using (\ref{eq:satval}) instead of the Boltzmann limit. The singularity of $N_{it}$ near \SI{0}{\kelvin} is also avoided. Since (\ref{eq:satval}) is independent of $V_{DS}$, the slightly higher $SS$ in Fig.\,\ref{fig:overview} at $V_{DS}=\SI{0.9}{\volt}$ compared to $V_{DS}=\SI{5}{\milli\volt}$ is a  high-field effect. 
\section{Conclusion}
An analytical expression for the saturating $SS(T)$ is derived from room down to sub-Kelvin temperature. When the thermal energy becomes smaller than the band-tail extension ($W_t$), the revised $SS(T)$ limit follows the temperature-independent $m(W_t/q)\ln 10$ rather than $m(k_BT/q)\ln10$. The revised limit demonstrates that a perfect MOS switch ($SS=0$) cannot be obtained in the presence of a band tail. The problem of extracting anomalously high interface-trap density at deep-cryogenic temperatures is solved by using $m(W_t/q)\ln 10$.  

\bibliographystyle{IEEEtran}
\bibliography{references}

\end{document}